\documentclass[a4paper,12pt]{article}
\usepackage[a4paper, left=2cm, right=2cm, top=2cm, bottom=2cm]{geometry}
\usepackage{graphicx}

\begin{document}

\begin{center}
\large{
 \textbf{Chemical order and crystallographic texture of FePd:Cu thin alloy films}}
\end{center}
\begin{center}
 Marcin Perzanowski, Yevhen Zabila, Michal Krupinski, Arkadiusz Zarzycki, Aleksander Polit, Marta Marszalek
\end{center}
\begin{center}
 \textit{Department of Materials Science, Institute of Nuclear Physics Polish Academy of Sciences, Radzikowskiego 152, 31-342 Krakow, Poland, Tel. (+48) 12 662 8145}
\end{center}

\begin{abstract}
FePd thin films have been recently considered as promising material for high-density magnetic storage devices. However, it is necessary to find a proper method of fabrication for the (001)-textured and chemically well-ordered alloy. In this paper, we present the detailed investigations of lattice parameters, chemical order degree, grain sizes and crystallographic texture, carried out on FePd alloys with $10$ at.\% of Cu addition. The initial [Cu($0.2$ nm)/Fe($0.9$ nm)/Pd($1.1$ nm)]$_{5}$ multilayers were thermally evaporated in an ultra-high vacuum on MgO(100), Si(100), Si(111) and Si(100) covered by $100$ nm thick layer of amorphous SiO$_{2}$. In order to obtain homogeneous FePd:Cu alloy, the multilayers were annealed in two different ways. First, the samples were rapidly annealed in nitrogen atmosphere at $600^{\circ}$C for $90$ seconds. Next, the long annealing in a high vacuum for $1$ hour at $700^{\circ}$C was done. This paper focuses on quantitative investigations of the chemical order degree and crystallographic texture of ternary FePd:Cu alloys deposited on four different substrates. In order to obtain both quantities we have taken a novel approach to consider the problem of dopant atoms located in the FePd structure. The studies of the structure were done using X-Ray Diffraction (XRD) performed with synchrotron radiation and pole figures measurements. We have found that the addition of Cu changes the FePd lattice parameters and lattice distortion. We have also shown, that using different substrates it is possible to obtain a FePd:Cu alloy with different chemical order and texture. Moreover, it was observed that texture category is substrate dependent.
\end{abstract}

\section{Introduction}
\label{introduction}

In recent years, a concept of perpendicular magnetic recording has been introduced in order to produce a new generation of high-density storage devices. L1$_{0}$-ordered intermetallic magnetic alloys, such as equiatomic FePd, FePt and CoPt, are the most promising candidates for this purpose, due to their uniaxial magnetic anisotropy in [001] direction \cite{Pir08JMM,Ter09JMM}. A possible industrial application of these materials strongly depends on the level of complication in producing (001)-textured and chemically well-ordered alloy.  

The perfect ordered single crystal bulk FePd alloy has AuCu-I structure type (space group P4/mmm) \cite{Vil96, Lau05SC}. In such a structure Fe atoms are placed in $1a$ and $1c$ Wyckoff sites ($(0,0,0)$ and $(\frac{1}{2},\frac{1}{2},0)$ positions in the unit cell), and Pd occupy $2e$ sites ($(\frac{1}{2},0,\frac{1}{2})$ and $(0,\frac{1}{2},\frac{1}{2})$ positions). The lattice parameters of bulk alloy are $a\!=\!3.855$~\AA~and $c\!=\!3.714$ \AA, which result in small lattice distortion in [001] crystallographic direction (axial ratio $c/a\!=\!0.963$). In order to obtain L1$_{0}$-ordered alloy it is necessary to tune the composition to approx. $50$ at.\% of both components \cite{Sun08PLA}.     

The most popular method of obtaining the L1$_{0}$-ordered and (001)-oriented FePd thin alloy films is the epitaxial growth. However, from the technical point of view, due to the specific growth conditions, this process is not convenient for mass production of the data storage devices. The other method for FePd alloy formation is the alternating deposition or codeposition of the constituent materials. Then, in order to produce alloys with L1$_{0}$ structure and proper magnetic properties, the as deposited systems are annealed by various post-deposition thermal processes. Crystallographic texture and chemical order could be induced by rapid thermal annealing \cite{Zen02APL, Yan03JAP}, conventional long time annealing \cite{Sat03JAP, Iss05SM}, or ion beam irradiation \cite{Sat03JAP, Iss05SM}. However, a lot of attention was also paid to investigate different approaches, such as annealing in high magnetic field \cite{Li04JMM, Cui10JAC}, combining ion irradiation and thermal treatment \cite{Kav06NI}, and growing films on heated substrates \cite{Car10JoP}.

It was found that the addition of various impurities to the L1$_{0}$ system can facilitate the formation of an ordered alloy, and many papers consider the dependence of chemical order, crystallite sizes, and magnetic properties on impurity additives. The formation of chemically ordered L1$_{0}$ phase of FePt alloys doped by Cu \cite{Pla02JAP, Wie04JMM, Yan06JAP}, Ag \cite{Zho03JAP}, and C \cite{Ko03APL} was observed. Also, the influence of Zr, W, Ti and TiO$_{2}$ or Ta$_{2}$O$_{5}$ additions on FePt alloy structure was reported in \cite{Kuo00JMM, Lee01APL, Che09JAP}, but there have been only few reports \cite{Nag06JAP, Nag06JAP-1, Kov07JAP, KanJKP09} on the doped L1$_{0}$-FePd thin films or nanoparticles so far. 

The definition of the chemical order parameter for binary alloys with L1$_{0}$ crystallographic structure can be found in \cite{War90}. This approach was successfully applied in studies concerning epitaxial growth of binary alloys \cite{Lis10N}, as well as in investigations of ion irradiation induced ordering in polycrystalline films \cite{Rav00APL} or L1$_{0}$-ordered nanoparticles \cite{Ron06AM}. However, the definition of chemical order parameter introduced in these papers is correct only for binary alloys. In case of ternary alloys we have to take into account the problem of site occupation by dopant atoms \cite{Mon06MSE, Nif06JCG}.

In this paper, we present the results of the studies on crystallographic texture and chemical order, carried out on FePd thin alloy films with $10$ at.\% of copper, investigated by X-Ray Diffraction and pole figure measurements. The initial Cu/Fe/Pd multilayers were deposited on four different substrates, and then transformed into the alloy by application of two thermal annealing procedures. Since the evaporation and annealing conditions were kept constant, it was possible to determine the influence both of substrate type as well as applied thermal treatment on the crystallographic properties of the FePd:Cu alloys. In order to obtain chemical order degree and quantitative information about crystallographic texture, we have introduced the new approach, which takes into account the problem of lattice site occupation by Cu addition atoms. It was found that these two quantities are dependent on the substrate type. Furthermore we will show qualitative analysis of the crystallographic texture which is also substrate-dependent.

\section{Experimental details}
\label{experimental_details}
The multilayer systems were prepared in an ultra-high vacuum chamber by thermal evaporation at working pressure in the range of $10^{-6}$ Pa. To provide $1\!:\!1$ stoichiometry between Fe and Pd atoms samples had [Cu($0.2$ nm)/Fe($0.9$ nm)/Pd($1.1$ nm)]$_{5}$ composition. The reference samples with composition [Fe($0.9$ nm)/Pd($1.1$ nm)]$_{5}$ were also prepared. The single layers were deposited sequentially and the thickness of each layer was monitored \textit{in-situ} using quartz crystal microbalance. Multilayer systems were evaporated on four substrates: polished MgO(100), Si(100), Si(111) and Si(100) covered by $100$ nm thick layer of amorphous SiO$_{2}$ (further denoted as Si(100)/SiO$_{2}$). Before the deposition, the substrates were ultrasonically cleaned in acetone and ethanol and rinsed in deionised water. Samples were deposited at room temperature, and the evaporation rates were $0.6$ nm/min for Fe and Pd layers and $0.2$ nm/min for Cu layers. The chemical composition was checked \textit{ex-situ} by Rutherford Backscattering (RBS). Nominal thicknesses of the single layers as well as total thickness of multilayer systems were confirmed \textit{ex-situ} by X-Ray Reflectivity (XRR) measurements. XRR measurements also showed, that on the top of the Si(100) and Si(111) substrates the layer of native silicon oxide with thickness of about $2$ nm was formed. All samples initially had size of $1$ cm per $1$ cm. After the RBS and XRR measurements the samples were cut into four pieces, and two of them were annealed with different annealing procedures.

In order to transform the initial multilayer system into the ordered alloy, the rapid thermal annealing method (further denoted as RTA) was used. The samples were annealed in an atmosphere of flowing nitrogen at $600^{\circ}$C for $90$ s, with a heating rate of $400^{\circ}$C/s. This procedure resulted in creation of the FePd:Cu alloy. It should be noted that due to the small size of sample (5 mm per 5 mm) the turbulences of flowing gas at the border of the samples could create the temperature gradient between the edge of the sample and its center. In addition to the RTA annealing the conventional long annealing (called LA) was applied at $700^{\circ}$C for $1$ hour in vacuum of $10^{-5}$ Pa. The heating rate was $15^{\circ}$C/min. 

The X-Ray Diffraction (XRD) experiments in $\Theta$/$2\Theta$ geometry were performed on CRISTAL beamline at SOLEIL synchrotron (France). The SOLEIL synchrotron storage ring has operated at $2.75$ GeV electron beam energy with a beam current of $400$ mA. At the CRISTAL beamline the in-vacuum U20 undulator insertion device was used as the source of X-ray radiation. The beamline was equipped with set of double-Si(111) single crystal flat monochromators with sagittal focusing. For the purpose of our measurements, X-ray radiation with energy of $10.000(1)$ keV was used, which corresponds to the wavelength of $0.123981(1)$ nm. At this wavelength, the longitudinal coherence length given by the $0.5 \lambda^{2}/\Delta\lambda$ expression was equal to $6$ $\mu$m. The horizontal and vertical size of the beamline spot was about $250$ $\mu$m per $100$ $\mu$m. For data collection, the proportional counter was used. All the measurements were carried out with angle $2\Theta$ changing from $15^{\circ}$ to $50^{\circ}$, the instrumental step of angle $2\Theta$ was $0.05^{\circ}$ with the counting time $1$ s per step.

The pole figures were measured using Panalytical X’Pert PRO laboratory diffractometer, equipped with three-circle Euler cradle. The Cu $K_{\alpha}$ radiation operated at $40$ kV, and $30$ mA was used. The optics of incident beam consisted of $1/4^{\circ}$ divergence slit, parabolic graded W/Si X-Ray mirror with an equatorial divergence of less than $0.05^{\circ}$, $0.04$ rad Soller slit collimator, and $4$ mm mask for restricting the axial width of incident beam. The diffracted signal was collected by solid state stripe X’Celerator detector with graphite monochromator working in receiving slit mode. The diffracted beam path was also equipped with $3.9$ mm high antiscatter slit and $0.04$ rad Soller slit collimator. The pole figures were measured using regular $5^{\circ}$ x $5^{\circ}$ grid in polar $\psi$ and azimuthal $\phi$ angles, with $\psi$ changed from $0^{\circ}$ to $80^{\circ}$ and $\phi$ from $0^{\circ}$ to $355^{\circ}$. The counting time was set for $90$ s per step. The X'Pert PRO diffractometer was also used for XRD measurements of FePd alloy on Si(100)/SiO$_{2}$. All the hardware settings were the same as for pole figure measurements. The $2\Theta$ range was choosen in a way corresponding to the range measured by synchrotron radiation. The stripe detector was working in scanning mode with active length of $2.122^{\circ}$. Instrumental step was $0.05^{\circ}$ and the counting time was $1600$ s per point. All the XRD and pole figures measurements were carried out at ambient conditions.

\section{Results and Discussion}
\label{results_and_discussion}
\begin{figure}
\centering
\includegraphics[width=8.5cm]{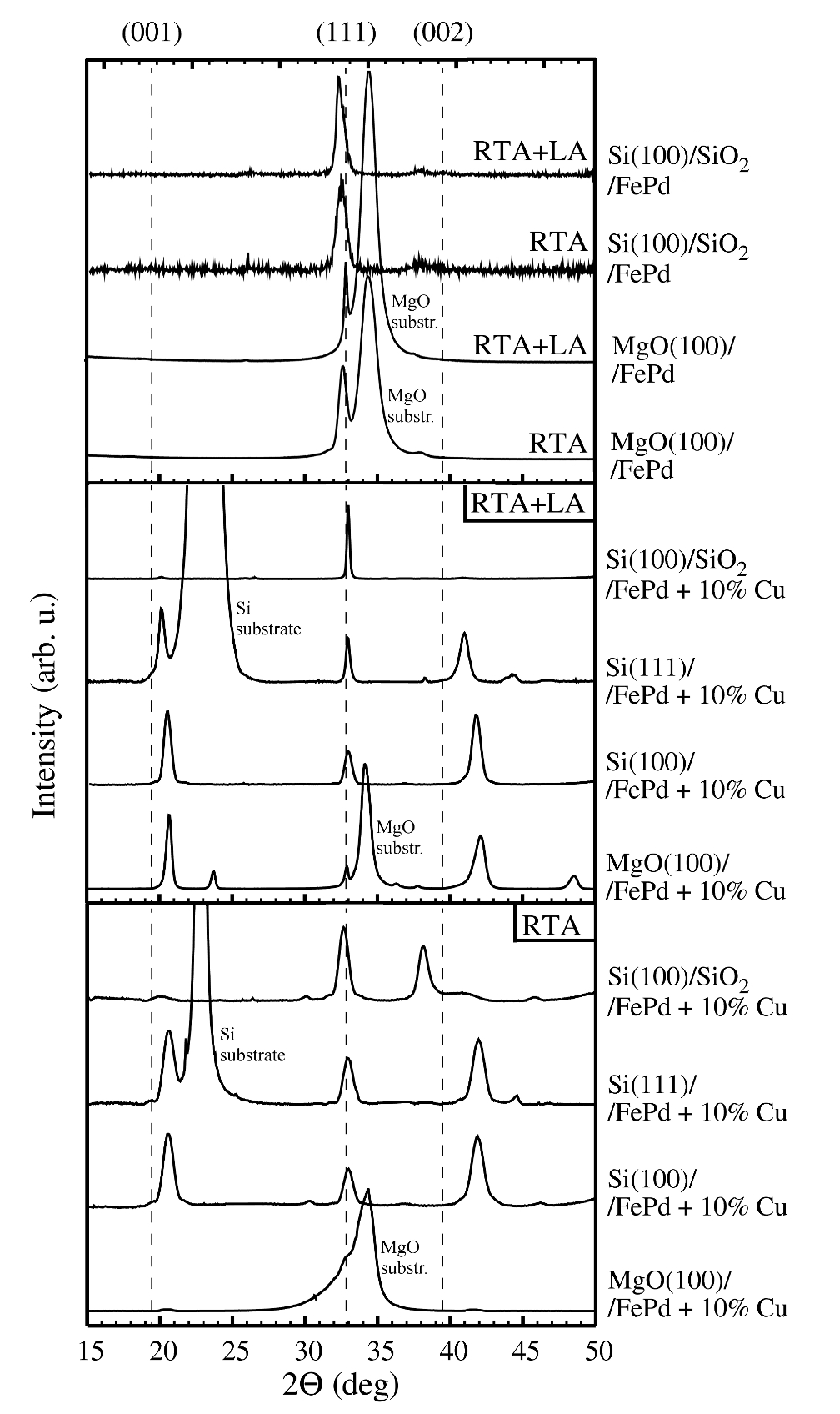}
\caption{XRD patterns of FePd alloys after RTA and RTA+LA (upper part) and FePd:Cu alloys after RTA (middle part) and RTA+LA annealing (lower part), deposited on MgO(100), Si(100), Si(111) and Si(100)/SiO$_{2}$ substrates. The positions of the (001), (111) and (002) reflections for the L1$_{0}$-ordered bulk FePd alloy are indicated with vertical dashed lines. The MgO and the Si substrate reflections are also indicated. All patterns were corrected for instrumental background and normalized to the intensity of the maximum intense sample reflection.}
\label{XRD_patterns}
\end{figure}

The XRD patterns, collected for FePd and FePd:Cu alloys after RTA and combined RTA+LA treatment, are presented in Fig.~\ref{XRD_patterns}. The positions of the (001), (111) and (002) reflections for L1$_{0}$-ordered bulk FePd alloy are marked in the figure with vertical dashed lines.

For FePd alloys only two reflections are observed in the patterns: the strong one at angle $2\Theta\!=\!33^{\circ}$ and very weak at $38^{\circ}$. In all patterns collected for FePd:Cu alloys, the set of three Bragg reflections at angles $2\Theta\!=\!20.5^{\circ}$, $33^{\circ}$ and $41.8^{\circ}$ are observed. The positions of the first and third peak are shifted towards higher angles in comparison the the bulk reflections, but the position of the second peak is the same as the position of bulk reflection. Additionally, for FePd:Cu alloy on Si(100)/SiO$_{2}$ substrate, after RTA, a strong reflection at $38.1^{\circ}$ appeared. It is readily observed that the intensity of the reflection at the same angular position in different samples is different and depends on sample type and annealing treatment. Changes of the integral intensities ratios between different reflections will be discussed in detail later. 

By comparing the bulk reflection positions and the positions of the observed peaks we can identify the reflections observed at angles $2\Theta$ $20.5^{\circ}$, $33^{\circ}$, and $41.8^{\circ}$ as coming from (001), (111) and (002) crystallographic planes of the L1$_{0}$-ordered alloys. Such result directly suggests the presence of only [111]-oriented crystallographic grains in FePd alloys and [001] and [111] orientations in case of FePd:Cu alloys. This is true for all cases, except of FePd:Cu alloy deposied on Si(100)/SiO$_{2}$ substrate, which exhibited additional (200) peak at $2\Theta\!=\!38.1^{\circ}$.

The presence of mostly [001] and [111]-oriented crystallites in case of FePd:Cu alloys is related to the impact of two main driving forces. The Cu addition, as well as the strain-inducing RTA annealing, introduce the grain orientation with c-axis perpendicular to the substrate plane. On the other hand, in metals and alloys with fcc or fct structure, the (111) plane is densly packed plane and has the smallest surface energy. Therefore, due to surface energy minimalization, the crystallites will tend to orient with this plane by being aligned parallel to surface. The lack of crystallites with [110] or [101] orientations is related to the fact that they have larger surface energy than (111), (001) or (100) planes and, consequently, are not energetically favoured. In case of FePd alloys the lack of Cu addition results in creation of [111]-oriented crystallographic grains according to the surface energy minimalization rule.

In order to obtain information about the values of lattice parameters $a$ and $c$, the integral intensities $I_{(hkl)}$ and full width at half maximum ($FWHM$) of the (001), (111) and (002) reflections, XRD data were fitted using the sum of pseudo-Voigt line profiles. Next, the values of lattice parameters $a$ and $c$ were calculated, based on the precise peak positions, from the standard definition of interplanar spacing $d_{\mathrm{(hkl)}}$.

\subsection{Lattice parameters and axial ratios}
\label{lattice_parameters_and_axial_ratios}
\begin{figure}[h]
\centering
\includegraphics[width=8.5cm]{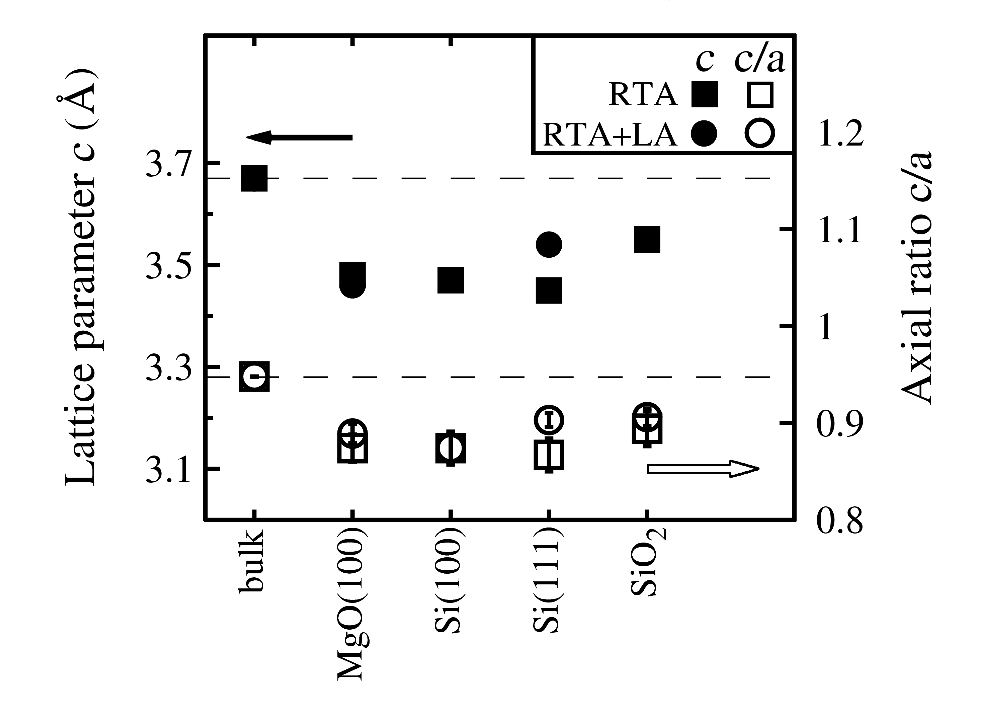}
\caption{The values of lattice parameter $c$ (full symbols, left scale) and $c/a$ axial ratios (open symbols, right scale) calculated from the XRD patterns. The values of lattice parameter $c$ for alloys on Si(100) and SiO$_{2}$ are the same after both annealing procedures. The bulk values of lattice parameter $c$ and axial ratio $c/a$ are marked with horizontal dashed lines (upper and lower lines, respectively).}
\label{lattice_parameters}
\end{figure}

For FePd alloys values of lattice parameters $a$ and $c$ and axial ratio $c/a$ are very close to the bulk values. The obtained values of lattice parameter $c$ and axial ratio $c/a$ for all FePd:Cu samples are shown in Fig.~\ref{lattice_parameters}. 
The shift of (001) and (002) peak positions towards higher angles is reflected by the smaller values of lattice paramater $c$ in comparison to bulk value. The lack of significant changes in (111) peak positions results in larger values for the lattice parameter $a$ than for the bulk material. The increase in values, together with simultaneous decrease of $c$ values, leads to larger tetragonal lattice distortion, visible as smaller values of axial ratios $c/a$ than for bulk material. The similar observation of lattice parameter changes was already reported in studies concerning Cu-doped FePt thin alloy films \cite{Pla02JAP, Wie04JMM, Yan06JAP}. It is worth noticing that the lattice parameter values change in such a way that the crystallographic $d_{\mathrm{(111)}}$-spacing remains always the same and is the same as the bulk value.  

For FePd alloys the $d_{(111)}$-spacing does not change since the lattice parameters have the same values as for bulk. In case of FePd:Cu alloys the lack of changes in the $d_{\mathrm{(111)}}$-spacing can be also explained by considering the energetic conditions. As it was mentioned before the (111) plane is a densely packed plane and has the lowest surface energy. The Cu addition to the FePd alloy causes reduction of the lattice parameter $c$ and disturbs the energetic equilibrium of the FePd L1$_{0}$ structure. On the other hand, the structure tends to energy minimum, which is related to the constant value of $d_{\mathrm{(111)}}$-spacing. Therefore, the lattice compression in [001] direction must be balanced by stretching in the [100] and [010] directions. The L1$_{0}$ (001) plane has the fourfold symmetry, so the stretching is equal in both directions and reflects in the expansion of the lattice parameter $a$.

\subsection{Long-range chemical order parameter}
\label{chemical_order_parameter}
Taking into account the definition given by Warren \cite{War90}, the long-range order parameter $S$ for alloys with L1$_{0}$ crystallographic structure can be expressed as corrected squared ratio between integral intensities of superstructure (001) and fundamental (002) Bragg reflections: 
\begin{equation}
\label{S_parameter}
S=\frac{\left| F_{\mathrm{(002)}} \right|}{\left| F_{\mathrm{(001)}} \right|} \sqrt{\frac{I_{\mathrm{(001)}} \left(G L P A_{\Theta/2\Theta} \right)_{\mathrm{(002)}}}{I_{\mathrm{(002)}} \left( G L P A_{\Theta/2\Theta} \right)_{\mathrm{(001)}}}}\mathrm{,}
\end{equation}

where $F_{\mathrm{(hkl)}}$ are structure factors, $I_{\mathrm{(hkl)}}$ are measured integral intensities, and $G$, $L$, $P$ and $A_{\Theta/2\Theta}$ are geometry, Lorentz, polarization and absorption factors, respectively (exact expressions are presented in \cite{Bir06}). The factors $G$, $L$ and $P$ depend only on angle $\Theta$, but factor $A_{\Theta/2\Theta}$ depends also on linear absorption coefficient $\mu$. The value of $\mu$ was calculated as the weighted sum of the absorption coefficients of Fe, Pd and Cu elements, with weights of their percentage mass contributions \cite{NIST}. The order parameter $S$ changes from $0$ for the lack of chemical order to $1$ for perfect chemical order in L1$_{0}$ alloy. Since the FePd alloys exhibit only (111) reflection the forthcoming analysis will be carried out only for FePd:Cu alloys.

The structure factors $F_{(001)}$ and $F_{(002)}$ in L1$_{0}$ structure are following:
\begin{equation}
\label{F_001}
F_{(001)}=\left( f_{1a}+f_{1c} \right) - \left( f_{2e'}+f_{2e''} \right),
\end{equation} 
and
\begin{equation}
\label{F_002}
F_{(002)}=\left( f_{1a}+f_{1c} \right) + \left( f_{2e'}+f_{2e''} \right),
\end{equation}
where $f_{1a}$, $f_{1c}$, $f_{2e'}$ and $f_{2e''}$ are atomic form factors associated with atoms placed in $1a$, $1c$ and two $2e$ Wyckoff sites in the unit cell. For perfect L1$_{0}$-ordered FePd alloy, the $1a$ and $1c$ positions are occupied by Fe atoms, and in $2e'$ and $2e''$ sites Pd atoms are placed. However, in case of ternary FePd:Cu alloy it is necessary to consider where copper atoms are located in the alloy structure. In our previous studies, about the local structure of the FePd:Cu alloy \cite{Kru11JAP, Pol11PRB}, it was shown that Cu atoms substitute both Fe and Pd sites. With the assumption of a random distribution of copper atoms in L1$_{0}$ structure the expressions for structure factors change to:
\begin{eqnarray}
\label{F_001_Cu}
F_{(001)}=\left[ \left( 0.9 f^{\mathrm{Fe}}_{1a}+0.1 f^{\mathrm{Cu}}_{1a} \right)+\left( 0.9 f^{\mathrm{Fe}}_{1c}+0.1 f^{\mathrm{Cu}}_{1c} \right) \right] \\
\nonumber - \left[ \left( 0.9 f^{\mathrm{Pd}}_{2e'}+0.1 f^{\mathrm{Cu}}_{2e'} \right)+\left( 0.9 f^{\mathrm{Pd}}_{2e''}+0.1 f^{\mathrm{Cu}}_{2e''} \right) \right],
\end{eqnarray} 
and
\begin{eqnarray}
\label{F_002_Cu}
F_{(002)}=\left[ \left( 0.9 f^{\mathrm{Fe}}_{1a}+0.1 f^{\mathrm{Cu}}_{1a} \right)+\left( 0.9 f^{\mathrm{Fe}}_{1c}+0.1 f^{\mathrm{Cu}}_{1c} \right) \right] \\
\nonumber + \left[ \left( 0.9 f^{\mathrm{Pd}}_{2e'}+0.1 f^{\mathrm{Cu}}_{2e'} \right)+\left( 0.9 f^{\mathrm{Pd}}_{2e''}+0.1 f^{\mathrm{Cu}}_{2e''} \right) \right].
\end{eqnarray}
The $f^{\mathrm{Fe}}$, $f^{\mathrm{Pd}}$ and $f^{\mathrm{Cu}}$ are atomic form factors of Fe, Pd and Cu atoms, respectively. 

\begin{figure}[h]
\centering
\includegraphics[width=8.5cm]{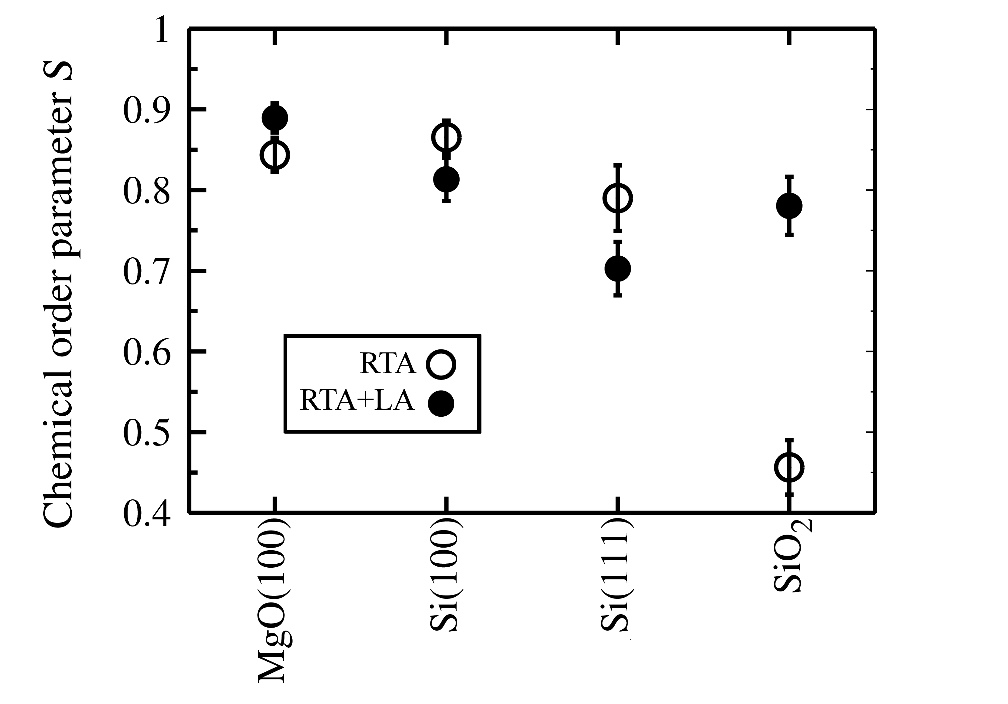}
\caption{The long-range order parameter $S$ for FePd:Cu alloys on various substrates after different annealing procedures.}
\label{s_parameter}
\end{figure}

Based on this model, the values of $S$ parameter were calculated, and the results are shown in Fig.~\ref{s_parameter}.
The largest $S$ value, close to $0.8$ -- $0.9$, was found for alloys deposited on MgO(100) and Si(100) substrates. Alloys deposited on Si(111) exhibited slightly smaller value of $S$. In case of these three substrates, the differences in chemical order, caused by different annealing procedures, are relatively small. However, it is worth noting, that in the case of Si(100) and Si(111) substrates, the addition of the second annealing stage led to a small decrease in the chemical order, in contrary to the sample on MgO(100) substrate, where another annealing stage did not change the parameter $S$ value. On the other hand, concerning the FePd:Cu alloy on the Si(100)/SiO$_{2}$ substrate, it is well seen, that RTA-annealed sample is not well-ordered ($S\!=\!0.45$). The significant increase of chemical order was observed after additional LA treatment ($S\!=\!0.8$). 

These results could be explained in such a way that, in case of single-crystal substrates, the chemical ordering process starts during RTA annealing with the intermixing of the ingredient layers in the multilayer system. Additional long annealing does not influence the chemical ordering process. For the FePd:Cu alloy on the Si(100)/SiO$_{2}$ substrate, the amorphous $100$ nm thick SiO$_{2}$ layer forces a two-step process of intermixing and ordering. The first annealing stage results in transformation from the multilayer system into alloy, then the second stage improves the chemical order.

\subsection{Preferred grain orientation}
\label{texture}
The presence of strong (111) and very weak (200) reflections in the XRD patterns for FePd alloys is the evidence of (111) texture presence. The lack of (001) and (002) reflections suggests that the two-stage annealling of FePd alloy does not create the texture with $c$-axis perpendicular to the substrate plane. In case of the FePd:Cu alloys the (111) and both (001) and (002) reflections are present in the patterns which is related to the existence of two texture components. For the quantitative analysis of the relation between these components let us consider two fundamental (002) and (111) Bragg reflections of L1$_{0}$ structure. The integral intensity of the (hkl) reflection can be expressed as \cite{Bir06}:
\begin{equation}
\label{integral_intensity}
I_{(hkl)} = X \left| F_{\mathrm{(hkl)}} \right|^{2} m_{\mathrm{(hkl)}} T_{\mathrm{(hkl)}} \left( G L P A_{\Theta/2\Theta} \right)_{\mathrm{(hkl)}},
\end{equation}
where $m_{\mathrm{(hkl)}}$ is the multiplicity of the reflection, $T_{\mathrm{(hkl)}}$ is the texture factor associated with the corresponding orientation, and $X$ is an instrumental factor related mostly to counting time and used setup (in present case is always constant due to the same experimental conditions). Factors $F_{\mathrm{(hkl)}}$, $G$, $L$, $P$ and $A_{\Theta/2\Theta}$ were already described.

In case of (002) and (111) reflections multiplicity $m$ equals to $2$ and $8$, respectively. The texture parameter $T_{\mathrm{(hkl)}}$ can be described as the fraction of the total sample volume, connected with (hkl)-oriented crystallographic grains. Considering the same model of Cu atom location in the unit cell as for chemical order parameter calculations, the expression for $F_{\mathrm{(111)}}$ is the same as expression \ref{F_002_Cu}. From expression \ref{integral_intensity} the ratio between texture factors $T_{(hkl)}$ can be expressed as:
\begin{equation}
\label{texture_parameters}
\frac{T_{\mathrm{(002)}}}{T_{\mathrm{(111)}}} = 4 \cdot \frac{I_{\mathrm{(002)}} \left| F_{\mathrm{(111)}} \right|^{2} \left (G L P A_{\Theta/2\Theta} \right)_{\mathrm{(111)}}}{I_{\mathrm{(111)}} \left| F_{\mathrm{(002)}} \right|^{2} \left (G L P A_{\Theta/2\Theta} \right)_{\mathrm{(002)}}}.
\end{equation}
The expression $T_{\mathrm{(002)}}/T_{\mathrm{(111)}}$ describes the ratio between total volumes of [001]- and [111]-oriented crystallites, and can have the following values:
\begin{itemize}
 \item for the lack of texture $T_{\mathrm{(002)}}/T_{\mathrm{(111)}}\!=\!1$,
 \item for [001] preferred orientation of grains $T_{\mathrm{(002)}}/T_{\mathrm{(111)}}\!>\!1$, 
 \item for [111] preferred orientation of grains $T_{\mathrm{(002)}}/T_{\mathrm{(111)}}\!<\!1$.
\end{itemize}
The described approach is valid only for two texture components in the sample. The presence of only \{001\} and \{111\} reflections in the measured XRD patterns allows to use this model. The exception is the RTA-annealed alloy deposited on Si(100)/SiO$_{2}$, where a strong (200) reflection appeared, however this case was treated as the others.

\begin{figure}[h]
\centering
\includegraphics[width=8.5cm]{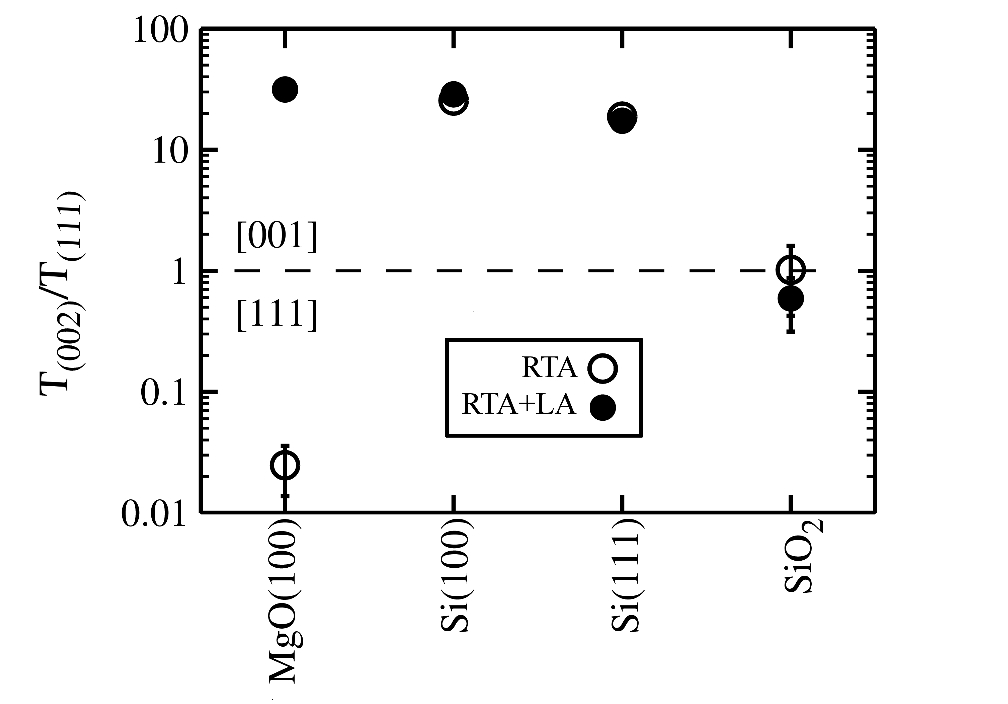}
\caption{Preferred grain orientation shown as texture factors ratio $T_{\mathrm{(002)}}/T_{\mathrm{(111)}}$. The value for the lack of texture is marked with horizontal dashed line.}
\label{preferred_orientation}
\end{figure}

The values of $T_{(002)}/T_{(111)}$ ratio for FePd:Cu alloys on different substrates are shown in Fig.~\ref{preferred_orientation}.
It is clearly seen that, for RTA-annealed FePd:Cu alloys deposited on Si(100) and Si(111) substrates, the strong (001) crystallographic texture appeared. Moreover, in case of these two substrates, the addition of the second annealing stage does not lead to a significant change in the $T_{\mathrm{(002)}}/T_{\mathrm{(111)}}$ ratio. A different effect can be observed in case of samples on MgO(100) substrate, where, after RTA, a strong (111) texture was recorded. The supplementary LA treatment led to a drastic reorientation of the crystallographic grains and strong (001) texture formation. After both annealing procedures, no well-defined crystallographic texture was found for Si(100)/SiO$_{2}$/FePd:Cu samples.

Let us discuss the problem of the (200) reflection presence in the pattern for RTA-annealed alloy on Si(100)/SiO$_{2}$ substrate. The thin film was deposited on the $100$ nm thick amorphous SiO$_{2}$ and we could expect, that the noncrystalline substrate might induce the formation of alloy without preferred grain orientation. This assumption is consistent with $T_{(002)}/T_{(111)}$ values, close to unity after both type of thermal treatment, which showed the same volume of [001] and [111]-oriented grains. The long annealing introduced the grain reorientation process which has evidence in the absence of [100]-oriented crystallites and lack of (200) reflection in the XRD pattern.

\begin{figure*}[t]
\centering
\includegraphics[width=17cm]{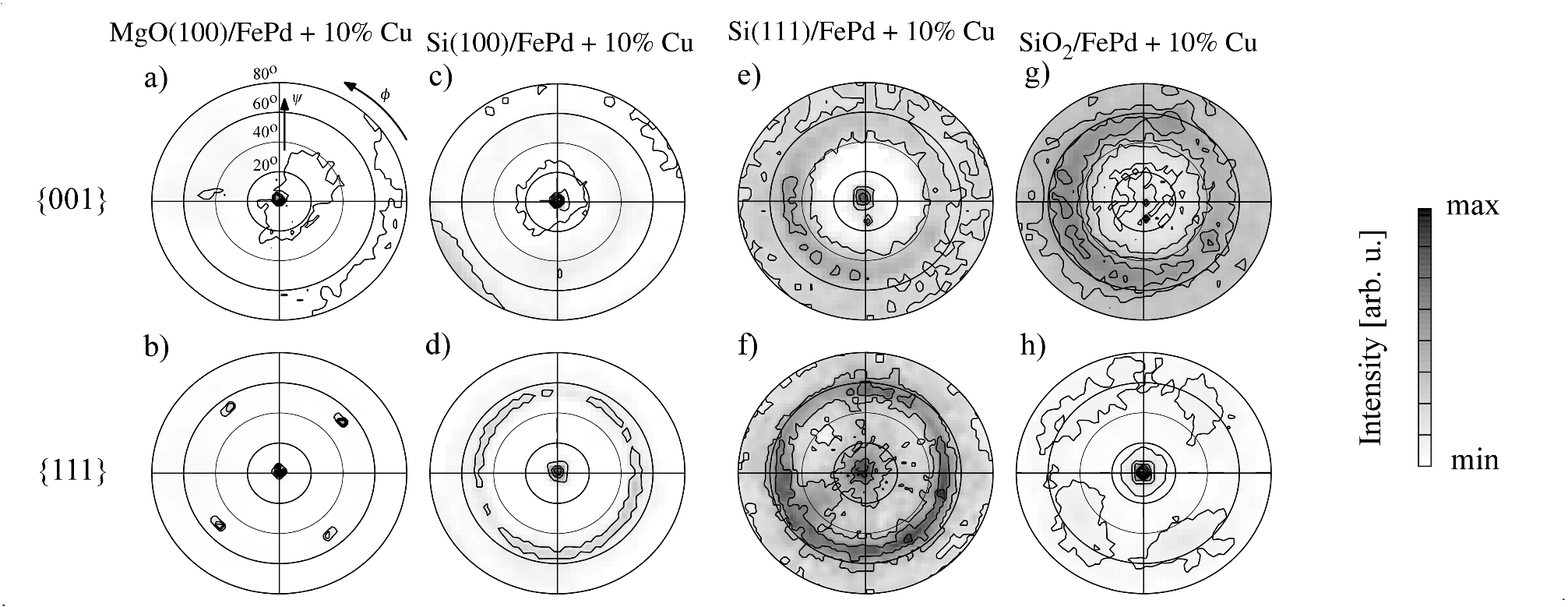}
\caption{The \{001\} (upper row) and \{111\} (lower row) pole figures for RTA+LA-annealed L1$_{0}$-ordered FePd:Cu samples on MgO(100) (a, b); Si(100) (c, d); Si(111) (e, f); and Si(100)/SiO$_{2}$ (g, h) substrates. For clarity of presentation pole figures were normalized between minimal and maximal signal value.}
\label{pole_figures}
\end{figure*}

In addition to preferred grain orientation determination, the \{001\} and \{111\} pole figure measurements were carried out for all RTA+LA-annealed samples. Pole figures were obtained using FePd:Cu L1$_{0}$ (001) and (111) reflections, and the results are shown in Fig.~\ref{pole_figures} (for precise determination of the peak positions the samples were re-measured in $\Theta$/$2\Theta$ geometry with Cu ${K_{\alpha}}$ radiation using laboratory diffractometer). The \{001\} pole figures for FePd:Cu alloys on MgO(100) and Si(100) substrates are nearly the same, with only one narrow peak in the centre of the patterns. However, the significant differences were found concerning the \{111\} pole figures, where two different types of signal were observed for $\psi$ about $52^{\circ}$. For the alloy on MgO(100), apart from the sharp central peak, four well-defined poles azimuthally separated by $90{^{\circ}}$ were recorded (Fig.~\ref{pole_figures}b). In the case of the Si(100)/FePd:Cu system, the sharp central pole is surrounded by a well-defined diffraction ring (Fig.~\ref{pole_figures}d). The observed signals are related to the [001]-oriented crystallographic grains in the alloy, and we can distinguish two types of crystallographic texture categories. For sample on MgO(100) the sharp (001) sheet texture is present, in contrast to the Si(100) substrate, where the sharp (001) fiber texture is observed with the [001] crystallographic direction as the fiber axis.

The \{001\} pole figure pattern observed for sample on Si(111) substrate consists of the central pole, surrounded by the $\psi$-dependent poles density distribution (Fig.~\ref{pole_figures}e). The distribution maximum is observed for $\psi$ about $52^{\circ}$, and suggests the presence of the (111) texture component with the [001] fiber axis. In case of the {111} pole figure, a similar pattern was measured, with a central pole and symmetrical poles distribution, with the maximum also at $\psi$ about $52^{\circ}$ (Fig.~\ref{pole_figures}f). The ring-shaped pole distribution is connected to the presence of the [001]-oriented grains with the [111] fiber axis.

For Si(100)/SiO$_{2}$/FePd:Cu system, the recorded \{001\} pole figure is similar to \{001\} pattern for the alloy on Si(111) substrate, indicating the (001) texture component with a [111] fiber axis. However, the central pole is not well-defined and the pole ring is more intense then for the Si(111)/FePd:Cu alloy. In the {111} pole figure, the only well-defined signal comes from the central pole (Fig.~\ref{pole_figures}g).

The pole figures show, that in case of annealed FePd:Cu thin alloy films the crystallographic texture category depends on the substrate type; the sheet texture was found for alloy on MgO(100), and silicon-based substrates provide fiber crystallographic ordering. Another aspect is that, for the MgO(100) and Si(100) substrates, the texture consists of two sharp (001) and (111) components. The results for alloys on the remaining substrates indicate blurred polar-dependent pole density distribution.

The appearance of [001]-oriented crystallites in FePd alloy is related mostly to two reasons: the Cu addition in the structure, and the fast and strain-inducing RTA annealing. However, the above mentioned reasons do not clarify the differences in the texture category for FePd:Cu alloys deposited on the various substrates. In the case of the FePd:Cu alloy on MgO(100), the single crystal substrate is perfectly ordered and oriented with the [100] direction perpendicular to the surface. Such an orientation, together with the symmetry of the MgO structure, causes the substrate surface to have a specific orientation defined by the [001] and [010] crystallographic directions. During the annealing process, the FePd:Cu crystallites tend to choose the direction, which minimizes the total energy of the system. Since the MgO substrate forces the two above mentioned directions, the (001) plane of L1$_{0}$ alloy is not oriented randomly, which is an explanation of the sheet texture of the FePd:Cu alloy on the MgO substrate. The described process is supported by the lattice misfit analysis. The MgO substrate has the same crystallographic symmetry as the L1$_{0}$ (001) plane. Taking into account the MgO and RTA+LA annealed FePd:Cu L1$_{0}$ (001) lattice parameters, the lattice misfit is about $7$\% (MgO lattice parameter ($a_{\mathrm{MgO}}\!=\!4.212$ \AA). This relatively small value can be the next indication,that FePd:Cu alloy crystallites tend to order along crystallographic directions of the MgO substrate.

In the case of the Si(100) and Si(111) substrates, the FePd:Cu alloy is formed on the disordered native Si-oxide layer. For such layer, there are no well-defined directions as in the case of the MgO substrate. During the annealing process, the FePd:Cu crystallites can orient randomly their (001) plane around the [001] direction, because, in the absence of substrate orientation, each of these configurations is equally favorable energetically. Therefore, the lack of defined crystallographic orientation of the substrate surface is the reason for the fiber texture of FePd:Cu alloys. For the same reason, the fiber texture is observed for FePd:Cu alloy on the Si(100)/SiO$_{2}$ substrate, but in this case the Si crystal was intentionally covered by amorphous Si-dioxide.

It is clearly seen in Fig.~\ref{pole_figures} that among the samples with fiber texture there are also significant differences in the sharpness of the orientation distribution. The reason of this effect is that the oxidation process of the Si (100) and (111) surfaces proceeds in a different way \cite{Iba82APA, Him88PRB} and results in the creation of the Si-oxide layers with slightly different properties. Moreover, the native oxide layers are not completely amorphous, which causes the FePd:Cu alloys deposited on these two substrates to reveal a larger $T_{\mathrm{(002)}}/T_{\mathrm{(111)}}$ ratio and better crystallographic order than those deposited on the Si(100)/SiO$_{2}$ substrate.

\subsection{Grain sizes}
\label{grain_sizes_a}
\begin{figure}[h]
\centering
\includegraphics[width=8.5cm]{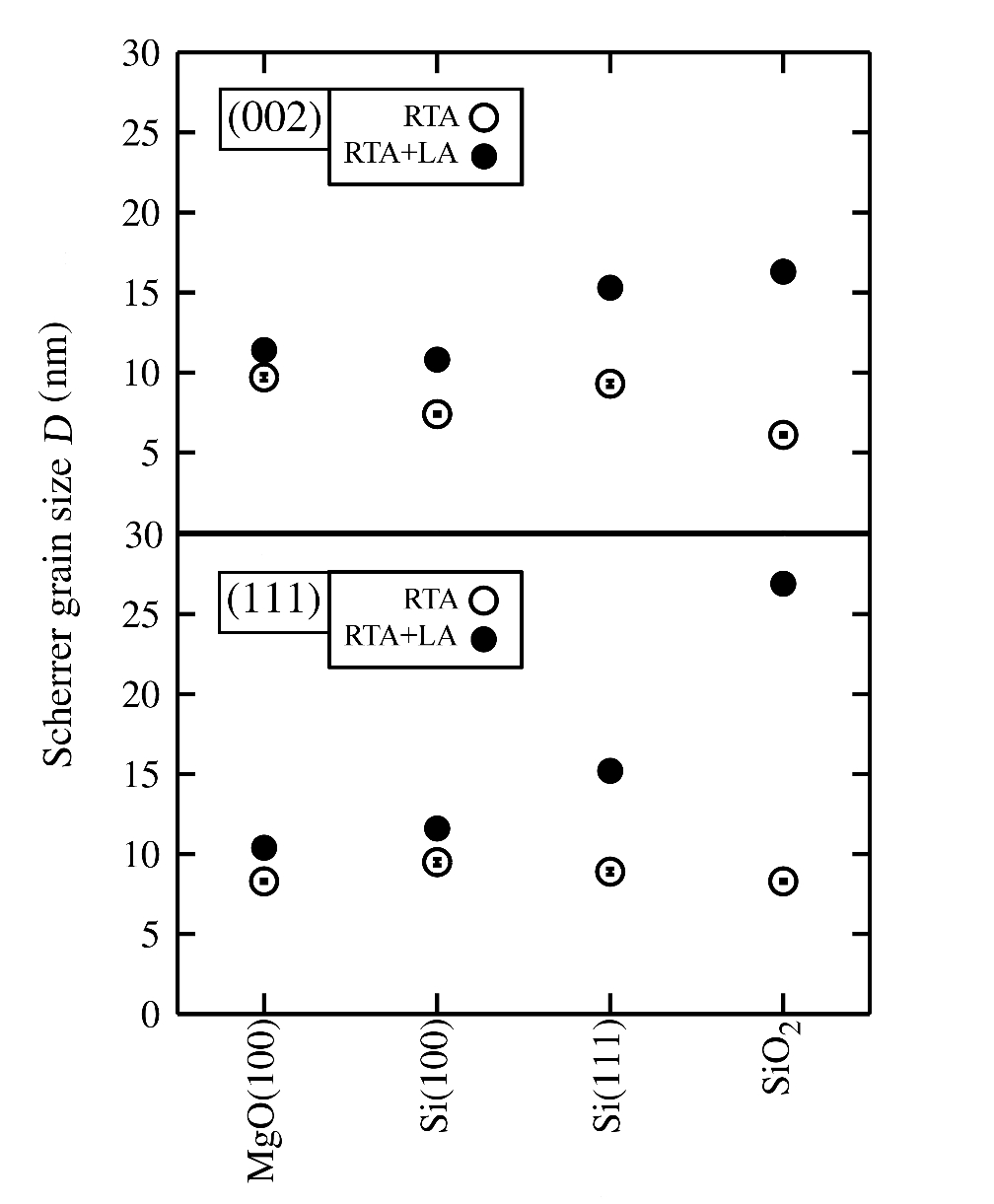}
\caption{Crystallographic grain sizes calculated using (002) (a) and (111) (b) reflections for the FePd:Cu alloys deposited on different substrates after RTA and RTA+LA annealing.}
\label{grain_sizes}
\end{figure}

In order to obtain information about the dependence of crystallographic grain size on the annealing treatment and substrate type the Scherrer equation was used:
\begin{equation}
\label{scherrer_equation}
D_{(hkl)} = \frac{K_{S} \lambda}{FWHM(2\Theta) \cos\Theta},
\end{equation}
where $K_{S}$ is a dimensionless Scherrer factor and its value is dependent on the crystallite shape \cite{Was04}, $\lambda$ is the radiation wavelength, $FWHM$ is a full width at half maximum for a (hkl) reflection at diffraction angle $\Theta$. For the purpose of these studies the value of $K_{S}\!=\!0.95$ was applied, as the most common used in similar studies. The calculations were carried out using (002) and (111) L1$_{0}$ fundamental reflections, indicating crystallite size along [001] and [111] crystallographic directions perpendicular to the substrate plane. The results of the calculations are shown in Fig.~\ref{grain_sizes}.

The main influence on the grain size has the type of annealing treatment. For both the [001] and [111] directions and for all substrates, crystallites obtained after RTA have sizes in the range of about $7$ nm to $10$ nm. After LA treatment, crystallite size increases to $10$ -- $16$ nm and $10$ -- $28$ nm, for the [001] and [111] directions, respectively. The grain sizes for samples after RTA+LA are also dependent on the substrate --- the smallest values were recorded for the alloy on MgO(100) and Si(100) substrates, and the largest for alloy on Si(100)/SiO$_{2}$. Moreover, in case of the Si(100)/SiO$_{2}$/FePd:Cu system, the significant differences between grain size along the [001] and [111] directions were found, and crystallites along the [111] axis are larger. 

It is seen that RTA annealing results in formation of nanocrystalline FePd alloy with grains of the size of a few nm at any type of substrate. Additional long time annealing increases the size of grains, and this effect is particularly strong for amorphous Si(100)/SiO$_{2}$ substrate. This might be due to the different character of the recrystallization process on an amorphous substrate, but for definite conclusion the microscopy measurements are necessary.

\section{Conclusions}
\label{conclusions}
We have investigated the crystallographic texture and the degree of the chemical order of the FePd:Cu thin alloy films, obtained by two different thermal treatments. The FePd thin alloy films were used as reference samples. The alloys were deposited on four different substrates, allowing the determination of the substrate influence on the structural properties of studied alloys.

Based on the detailed XRD studies and pole figure measurements, we have found that the Cu addition changes lattice parameters. The lattice parameter $c$ decreases with simultaneous increase of lattice parameter $a$, leading to the larger lattice distortion, but the crystallographic $d_{\mathrm{(111)}}$-spacing stays always the same. The calculations of the chemical order parameter $S$ were carried out with the assumption of the random distribution of copper atoms in the FePd L1$_{0}$ lattice sites. The largest values of parameter $S$ were obtained for the FePd:Cu alloys on MgO(100) and Si(100) substrates. There is no relationship between chemical order and preferential orientations of grains. We have observed that the degree of ordering is very similar for samples deposited on MgO and Si(100), most likely due to the same symmetry of the substrate surfaces as the symmetry of L1$_{0}$ phase. In addition to it, the strong contribution of [001] texture was observed for all alloys deposited on single crystalline substrates although the different sheet and fiber characters of texture were observed for MgO and Si, respectively. This can be related to epitaxial-like growth of films on MgO, in contrary to nonepitaxial growth on Si. For alloys on the Si-based substrates, also the differences in the crystallographic texture were found. The alloy deposited on Si(100) exhibits two well defined [001] and [111] components, whereas the signals for alloys deposited on Si(111) and Si(100)/SiO$_{2}$ were blurred, indicating the degradation of crystallographic orientation distribution. As could be expected the amorphous Si/SiO$_{2}$ surface did not induced any preferential orientation of films. 

The Scherrer analysis showed that the annealing treatment has a strong influence on crystallite sizes. In all samples, after RTA annealing, grains are the size of a few nanometers. The additional long annealing leads to grain growth, except for the alloy deposited on MgO substrate. The largest change in grain size was found for the alloy on the Si(100)/SiO$_{2}$ substrate, where the grain growth mostly in [111] direction was recorded. According to the presented studies, the MgO(100) and Si(100) substrates were found to be the most suitable for preparation of the chemically ordered and (001)-textured polycrystalline FePd:Cu thin alloy films. 

\section*{Acknowledgements}
The investigations were partially supported by DAAD-Polish Ministry of Science and Higher Education program of Polish-German cooperation under the contract 329/N-DAAD/2008/0 and Polish Ministry of Science and Higher Education project N$^{\circ}$ N N507 500338. The support of Dr. Erik Elkaim from SOLEIL Synchrotron by working on proposal N$^{\circ}$ 20090217 and Dr. Denys Makarov from Institute for Integrative Nanosciences IFW Dresden is also acknowledged.

\end{document}